\documentclass[paper=a4,abstracton,numbers=noenddot]{scrartcl}

\usepackage[english]{babel}  
\usepackage{hyperref}
\usepackage[intlimits]{amsmath}
\usepackage{amsthm}
\usepackage{amssymb} 
\usepackage{units} 
\usepackage{amsfonts,bbm}
\allowdisplaybreaks[1]
\usepackage[numbers]{natbib}
\usepackage{graphicx}
\usepackage[space]{grffile} % + graphicx: allow figure names with spaces
\usepackage[T1]{fontenc}
\usepackage{lmodern}  % latin modern fonts - added to resolve visualization issues with fontenc			
\usepackage[scaled=.92]{helvet}
\usepackage{courier}

\usepackage[a4paper,left=2.5cm,right=2.5cm,top=3cm,bottom=2.5cm]{geometry}
\usepackage{authblk}
\usepackage{mathtools}
\usepackage{caption}
\newcommand{\OfOrder}{\mathcal{O}}
\newcommand\pt[1]{\boldsymbol{#1}}

\newcommand\pu{p}			  % NURBS degree on u
\newcommand\pdim{d}			% Parametric dimension
\newcommand{\TrimCurve}{\Gamma}     % trimming curve
\newcommand{\domain}{\Omega}
\newcommand{\patchdomain}{\domain^{\textnormal{v}}}
\newcommand{\supportdomainReg}{\mathcal{S}^{\textnormal{r}}}
\newcommand{\supportdomain}{\mathcal{S}^{\textnormal{v}}}
\newcommand{\supportdomainCut}{\mathcal{S}^{\textnormal{c}}}
\newcommand\indexA{i} 		% first normal index
\newcommand{\Bspline}{B} 		% symbol for B-spline
\newcommand\uu{\xi} 			% Local NURBS coordinates
\newcommand\UVsurf{\pt{\uu}} 	% Local NURBS coordinates
\newcommand{\discCoeff}{{\uu}^{\textnormal{disc}}} 
\newcommand{\R}{\mathbb{R}}

\DeclareMathOperator{\supp}{supp}
\newcommand{\Supp}[1]{\supp\{ #1\}}

\title{Fast formation and assembly for spline-based 3D fictitious domain methods}

\author{Benjamin Marussig}

\affil{Institute of Applied Mechanics, Graz Center of Computational Engineering (GCCE), Graz University of Technology, Technikerstra\ss e 4/II, 8010 Graz, Austria}
\date{}                     %% if you don't need date to appear
\begin{document}

\maketitle

\begin{abstract}
  Standard finite element methods employ an element-wise assembly strategy. 
  The element's contribution to the system matrix is formed by a loop over quadrature points. 
  This concept is also used in fictitious domain methods, which perform simulations on a simple tensor-product background mesh cut by a boundary representation that defines the domain of interest. 
  
  Considering such $d$-dimensional background meshes based on splines of degree $p$ with maximal smoothness, $C^{p-1}$, the cost of setting up the system matrix is $\OfOrder\left(p^{3d}\right)$ per degree of freedom. 
  Alternative assembly and formation techniques can significantly reduce this cost. 
  In particular, the combination of (1) sum factorization, (2) weighted quadrature, and (3) row-based assembly yields a cost of $\OfOrder\left(p^{d+1}\right)$ for non-cut background meshes. 
  However, applying this fast approach to cut background meshes is an open challenge since they do not have a tensor-product structure.
  
  This work presents techniques that allow the treatment of cut background meshes and thus the application of fast formation and assembly to fictitious domain methods. 
  First, a discontinuous version of weighted quadrature is presented, which introduces a discontinuity into a cut test function's support. 
  The cut region can be treated separately from the non-cut counterpart; the latter can be assembled by the fast concepts. 
  A three-dimensional example investigates the accuracy and efficiency of the proposed concept and demonstrates its speed-up compared to conventional formation and assembly.
\end{abstract}

\section{Introduction}

\begin{figure}
  \hfil
  \includegraphics[width=0.3\textwidth]{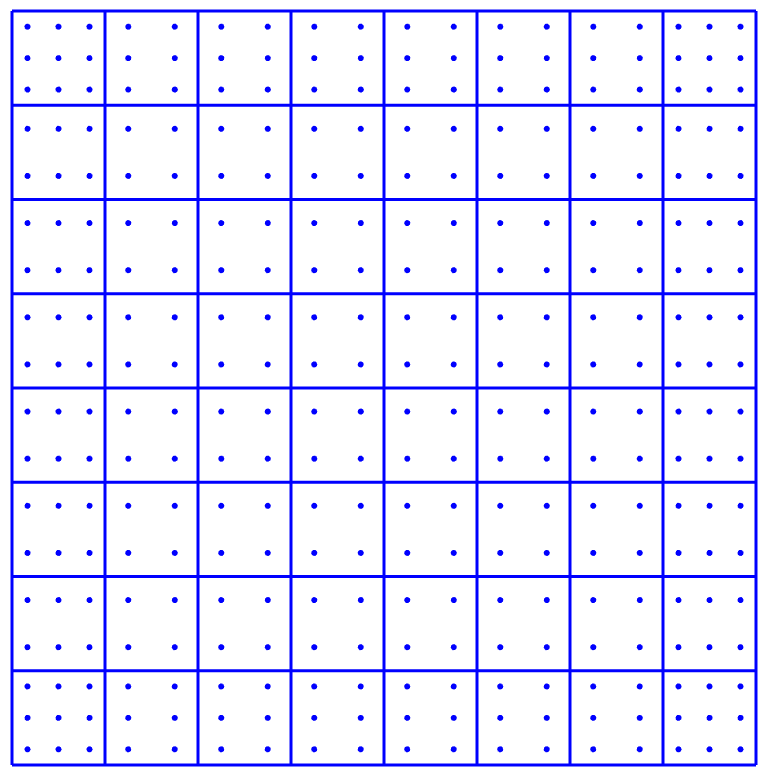}~a)
  \hfil
  \includegraphics[width=0.3\textwidth]{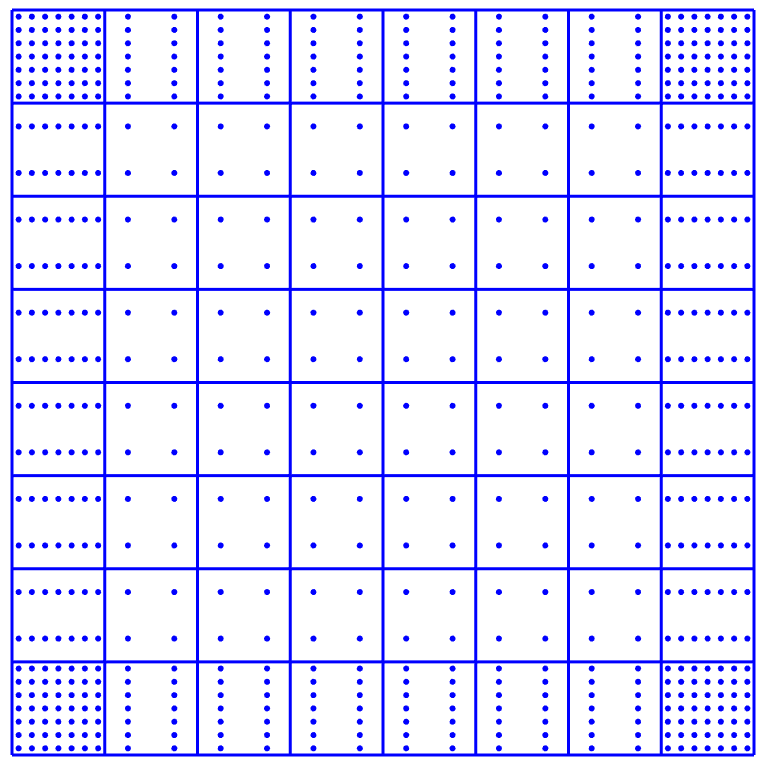}~b)
  \hfil
  \caption{Weighted quadrature points for two 2d background meshes defined by B-splines with maximal smoothness and the same number of elements but different polynomial degree: \textbf{a} bi-degree $\pu=2$ and \textbf{b} bi-degree $\pu=6$.}
  \label{fig:1}
\end{figure}

%% spline-based background meshes 
Fictitious domain methods introduce a regular background mesh for performing numerical simulations. 
B-splines can represent such meshes effectively, allowing a straightforward definition of high-order parametrizations with complete control over the regularity between the mesh elements. 
%% draw back of standard approach
During the simulation, an element-wise assembly procedure is state-of-the-art at the core of standard finite element codes.
Yet, the corresponding cost for setting up the system matrix for a $C^{\pu-1}$-continuous $\pdim$-dimensional tensor product B-spline basis is $\OfOrder\left(\pu^{3\pdim}\right)$ per degree of freedom \cite{Calabro2017a}. 
Thus, high-order analysis becomes computationally expensive using conventional matrix formation and assembly.
%
%% key ingredients and references
The fast formation and assembly concept reduces the computational cost to $\OfOrder\left(\pu^{\pdim+1}\right)$ by using the following essential ingredients:
\begin{itemize}
  \item \textbf{Sum factorization} \cite{Antolin2015a,Orszag1980a,Bressan2019a}: The concept exploits the tensor product structure of B-splines by rearranging computations of $d$-dimensional integrals such that only 1-dimensional integrals occur.
  \item \textbf{Weighted quadrature} \cite{Calabro2017a,Hiemstra2017a,Giannelli2022a}: A quadrature rule is set up for each test function by incorporating the test function into the quadrature weights. By exploiting the smoothness of B-splines, the resulting number of quadrature points is independent of the degree $\pu$ for splines with maximal smoothness, i.e., $C^{\pu-1}$, except at the boundary where B-splines (with open knot vectors) are discontinuous, i.e., $C^{-1}$.
  For multi-variate B-splines, quadrature rules are computed for each parametric direction; the tensor product of these points determines the overall layout of quadrature points, as shown in Fig.~\ref{fig:1}.
  \item \textbf{Row assembly}: The assembly employs a loop over the test functions, i.e., the rows of the system matrix, instead of an iteration over elements.  
\end{itemize}

%% impact of fast formation and assembly 
The effect of combining these ingredients is summarized in Tables \ref{tab:1} and \ref{tab:2} in terms of the required floating point operations (FLOPS) for setting up a 3D mass matrix. 
For details the interested reader is referred to \cite{Hiemstra2017a}.
The tables' columns distinguish the element formation strategy (loop over quadrature points or sum factorization), and the rows denote the assembly scheme (loop over elements or rows). 
The tables themselves refer to different quadrature rules; Table~\ref{tab:1} considers quadrature rules that require $p^d$ points per element (e.g., Gauss quadrature), whereas Table \ref{tab:2} assesses weighted quadrature.
In other words, Table~\ref{tab:1} (row 1/column 1) represents the standard finite element procedure, while Table \ref{tab:2} (row 2/column 2) describes the proposed scheme.
Note that the combination of sum factorization, weighted quadrature, and row assembly is vital for the resulting significant drop in the computational cost.
\begin{table}[h]
  \centering
  \begin{tabular}{@{}l|ll@{}}
    \hline
    ${\text{Assembly}} \backslash {\text{Formation}}$	&  Quadrature loop	& Sum factorization		\\
    \hline
    Element loop 		& $c\cdot p^9 $ & $c_1\cdot p^5 + c_2\cdot p^6 + c_3\cdot p^7 $ \\ %\cline{2-4}
    Row loop & $c\cdot p^9 $ & $c_1\cdot p^7 + c_2\cdot p^6 + c_3\cdot p^5 $ \\ %\hline
    \hline
  \end{tabular}
  \caption{FLOPS for setting up a 3D mass matrix with a quadrature rule using $\OfOrder\left(\pu^{3}\right)$ points per element such as Gauss quadrature \cite{Hiemstra2017a}.}
  \label{tab:1}
\end{table}
\begin{table}[h]
  \centering
  \begin{tabular}{@{}l|ll@{}}
    \hline
    ${\text{Assembly}} \backslash {\text{Formation}}$	&  Quadrature loop	& Sum factorization		\\ 
    \hline
    Element loop 		& $c\cdot p^6 $ & $c_1\cdot p^2 + c_2\cdot p^4 + c_3\cdot p^6 $ \\ %\cline{2-4}
    Row loop 	& $c\cdot p^6 $ & $c_1\cdot p^4 + c_2\cdot p^4 + c_3\cdot p^4 $ \\ %\hline
    \hline
  \end{tabular}
  \caption{FLOPS for setting up a 3D mass matrix with a quadrature rule using $\OfOrder\left(1\right)$ points per element such as weighted quadrature \cite{Hiemstra2017a}.}
  \label{tab:2}
\end{table}
%

%% fictitious domain methods
In fictitious domain methods, an interface $\TrimCurve$ embedded in the background mesh specifies the domain of interest. Unfortunately, the intersection with $\TrimCurve$ destroys properties required for sum factorization and weighted quadrature. 
A discontinuous version of weighted quadrature has been proposed in \cite{Marussig2022inproc} to address this issue for two-dimensional domains. 
%% research gap 3D + Gauss pts for base-layout
In this work, this approach is extended to the three-dimensional setting and further improved by reducing the number of quadrature point evaluations needed.

\section{Fast assembly and formation for fictitious domains}

Without a loss of generality, it is assumed that the identity map provides the transformation from the parameter space to the background mesh. This assumption simplifies the following descriptions since the background mesh and the parameter space coincide, and we do not have to distinguish between the interface $\TrimCurve$ in the parametric and physics domain.

\subsection{Function types of cut background meshes}

\begin{figure}[h]
  \centering
  \includegraphics[scale=1.0]{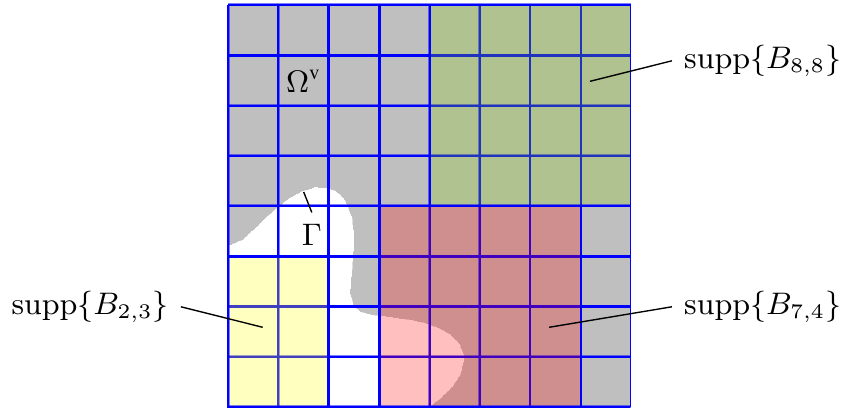}
  \caption{Cubic bi-variate basis with the interface $\TrimCurve$ specifying the valid domain $\patchdomain$ (gray). The resulting B-splines types are interior (green), cut (red), or exterior (yellow) based on the overlap of the support, $\Supp{ \Bspline_{\indexA_1,\indexA_2} }$, with $\patchdomain$.}
  \label{fig:basisFunctionTypes}
\end{figure}

The presence of $\TrimCurve$  splits the background mesh into an interior and exterior domain, where the former specifies the valid domain of interest $\patchdomain$. 
Furthermore, it divides the basis functions into three different types.
That is, each B-spline $\Bspline_{{\boldsymbol{\indexA}}}$ of the basis belongs to one of the following categories based on the overlap of its support with $\patchdomain$, i.e., $\supportdomain_{\boldsymbol{\indexA}} \coloneqq  \Supp{ \Bspline_{{\boldsymbol{\indexA}}} }  \cap {\overline{\patchdomain}}$:
\begin{itemize}
    \item \emph{Exterior} if $\supportdomain_{\boldsymbol{\indexA}} = \emptyset$, 
    \item \emph{Interior} if $\supportdomain_{\boldsymbol{\indexA}} = \Supp{ \Bspline_{{\boldsymbol{\indexA}}} }$, 
    \item \emph{Cut} if  {$0 < \left|\supportdomain_{\boldsymbol{\indexA}}\right| <\left| \Supp{ \Bspline_{{\boldsymbol{\indexA}}} }\right|$,} 
\end{itemize}
where $\left|\cdot\right|$ denotes the Lebesgue measure in $\R^\pdim$.
Fig.~\ref{fig:basisFunctionTypes} shows examples of these different types.
Exterior B-splines can be neglected from the system of equations, and interior ones can be treated directly by the fast formation and assembly concept.
The integration of cut functions, however, needs special considerations detailed in the next section.

\subsection{Discontinuous weighted quadrature}

\begin{figure}%[b!]
  \centering
  \includegraphics[scale=0.82]{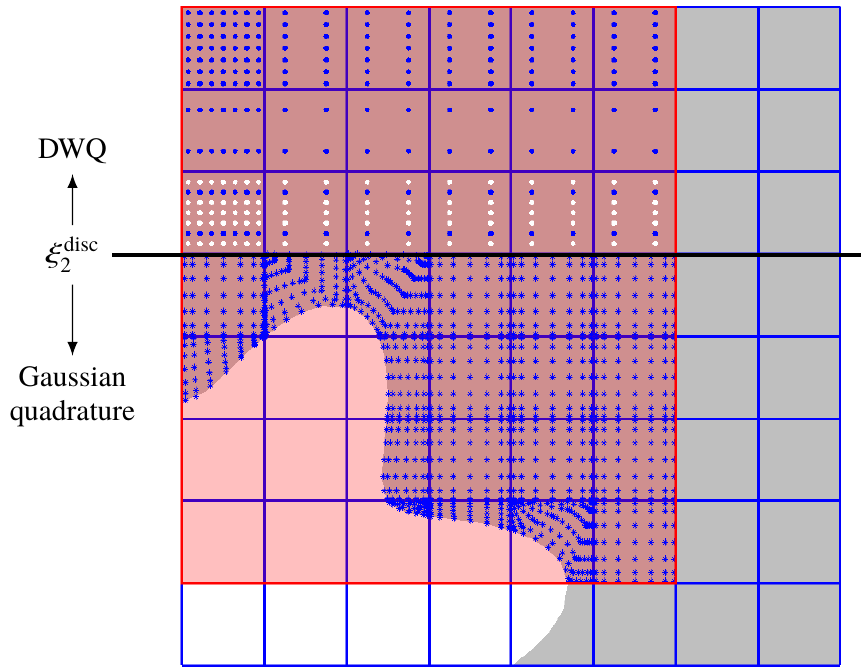}
  \caption{All quadrature points of a cut B-spline of bi-degree~6: above the artificial discontinuity $\discCoeff_2$ discontinuous weighted quadrature (DWQ) is employed, while Gauss points are used below. The white dots mark the nested quadrature points added during the construction.}
  \label{fig:TrimmedSpaceDiscWQExampleB}
\end{figure}

%% split of the support in cut and interior elements
First, the domain $\supportdomain_{\boldsymbol{\indexA}}$ is split into a \emph{regular} part $\supportdomainReg_{\boldsymbol{\indexA}}$, which follows the tensor product structure (at least on the element-level), and a \emph{cut} part $\supportdomainCut_{\boldsymbol{\indexA}}$, which consists of all elements cut by the interface.
The integral over a cut basis function can be written as
\begin{align}
    \label{eq:domainSplitting}
    \int_{\supportdomain_{\boldsymbol{\indexA}}}    \Bspline_{\boldsymbol{\indexA}}(\UVsurf) d\UVsurf = 
    \int_{\supportdomainReg_{\boldsymbol{\indexA}}} \Bspline_{\boldsymbol{\indexA}}(\UVsurf) d\UVsurf + 
    \int_{\supportdomainCut_{\boldsymbol{\indexA}}} \Bspline_{\boldsymbol{\indexA}}(\UVsurf) d\UVsurf. 
\end{align}
The numerical integration of ${\supportdomainCut_{\boldsymbol{\indexA}}}$ employs a standard element-wise procedure detailed in \cite{Fries2015a}.
The following focuses on treating the remaining regular part $\supportdomainReg_{\boldsymbol{\indexA}}$ using discontinuities weighted quadrature (DWQ).
The basic steps are (see \cite{Marussig2022inproc} for details on each step):
\begin{enumerate}
  \item For each parametric direction $d$, set up the standard weighted quadrature rules by defining the location of the weighted quadrature points and computing their weights. 
  \item Find the knot value $\discCoeff_d$ for each cut B-spline $\Bspline_{\boldsymbol{\indexA}}$ that maximizes the number of interior elements arranged in a tensor product structure. Fig.~\ref{fig:TrimmedSpaceDiscWQExampleB} illustrates the situation for a single $\Bspline_{\boldsymbol{\indexA}}$ in a two-dimensional domain, where $\discCoeff_2$ becomes a parametric line splitting $\Supp{ \Bspline_{{\boldsymbol{\indexA}}} }$ into two parts. The upper can be integrated by DWQ rules constructed in the subsequent steps.
  \item For each $\discCoeff_d$, an artificial $C^{-1}$ discontinuity at $\discCoeff_d$ is introduced by performing knot insertion into the corresponding knot vector, and the related subdivision matrix $\mathbf{S}$ is stored.
  \item Due to the introduced discontinuity, the minimal number of weighted quadrature points increases. Hence, we need to add new nested quadrature points. In Fig.~\ref{fig:TrimmedSpaceDiscWQExampleB}, these are the white dots above $\discCoeff_2$. 
  \item Compute the weighted quadrature weights $\tilde{w}$ for the refined univariate basis functions.
  \item Multiple these weights $\tilde{w}$ by $\mathbf{S}^{\textnormal{T}}$ to obtain the weights $w$ for the initial univariate basis functions. These are the weights of the DWQ rule corresponding to $\discCoeff_d$. 
\end{enumerate}
In contrast to standard weighted quadrature rules, the DWQ-weight-values can be set to zero on one side of the artificial discontinuity $\discCoeff_d$ without affecting the numerical integration result on the other side. Each cut B-spline has one associated DWQ rule and the other parametric directions are integrated with the standard weighted quadrature rules.
When the DWQ procedure is applied to three-dimensional domains, each $\discCoeff_d$ introduces a plane that splits the cut basis function's support. Hence, the detection of $\discCoeff_d$ is more involved. However, the remaining steps stay the same.

Note that the user can define the layout of the weighted quadrature points. This work proposes using the location of the standard element-wise Gauss quadrature points as a superset for choosing the initial and nested weighted quadrature points.
This arrangement allows the reuse of evaluations at the quadrature points independent of the rule applied. In addition, the computation of DWQ rules can be omitted if the number of nested and initial quadrature points equals the number of Gauss points; in such cases, the Gauss weights can be used directly.

\section{Numerical investigations}

\subsection{Test setting}
\begin{figure}[b!]
  \centering
  \includegraphics[width=0.25\textwidth]{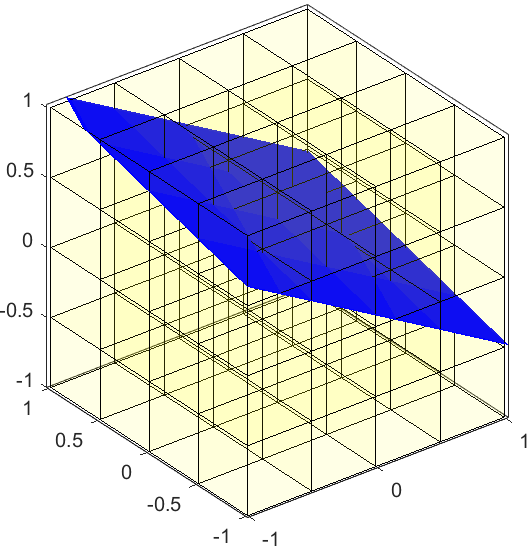}
  \caption{Geometric description of the numerical example: The background mesh $\domain = [-1,1]^3$ is parametrized by the degree $p$ and the number of elements $h$ per direction ($h=4$ in the example shown). The point $(0.1, 0.2, 0.3)^T$ and normal vector $(0.5,-0.2, 0.9)^T$ define the interface $\TrimCurve$ illustrated in blue, and the part below of this inclined plane determines the domain of interest $\patchdomain$.}
  \label{fig:exGeometry}
\end{figure}

%% problem set up
The numerical experiments focus on the formation of the mass matrix for the three-dimensional fictitious domains defined in Fig.~\ref{fig:exGeometry} 
and the subsequent $L^2$-projection onto the target function $f = \sin(2xz) \cos(3yz).$

Different background meshes are constructed by varying the degree $p=\{2,\dots,6\}$ and the number of elements per dimension $h=\{4,8,16,32,64\}$.
%% quantification
The relative $L^2$-error norm $\| \epsilon_{rel}\|_{L^2}$ of the resulting approximation to the target function quantifies the accuracy. 
In addition, the timings for setting up the corresponding mass matrices assess the efficiency. All routines have been implemented in an in-house MATLAB\textregistered~code, which does not utilize parallelization capabilities, and timings have been measured with MATLAB's tic-toc command.   
%% further approaches used
The mass matrix's conditioning affects the results' quality and may suffer due to the presents of cut elements \cite{marussig2017a}. 
Therefore, the extended B-spline concept is employed for all simulations to guarantee well-conditioned mass matrices. 
The interested reader is referred to \cite{Hoellig2003b,marussig2016a,marussig2018a} for details on this approach. 

%

%% comparisons
In order to assess performance, the following assembly and formation concepts are compared:
\begin{itemize}
  \item \textbf{Standard Gauss (Ref)}: Gauss rules and quadrature point loops are used for all interior elements. This scheme serves as reference to the conventional simulation paradigm. 
  \item \textbf{Hybrid Gauss}: Weighted quadrature evaluates interior test functions, while Gauss quadrature with sum factorization is used for the regular support $\supportdomainReg$ of cut test functions.
  \item \textbf{Discontinuous weighted quadrature (DWQ)}: Weighted quadrature evaluates interior test functions, while DWQ is used for the regular support $\supportdomainReg$ of cut test functions.
\end{itemize}
In all cases, the numerical integration of cut elements utilizes the approach detailed in \cite{Fries2015a} which decomposes cut elements into sub-elements that then employ standard Gauss quadrature rules.

\subsection{Results}

\begin{figure}[h]
  \centering
  \includegraphics[scale=1.0]{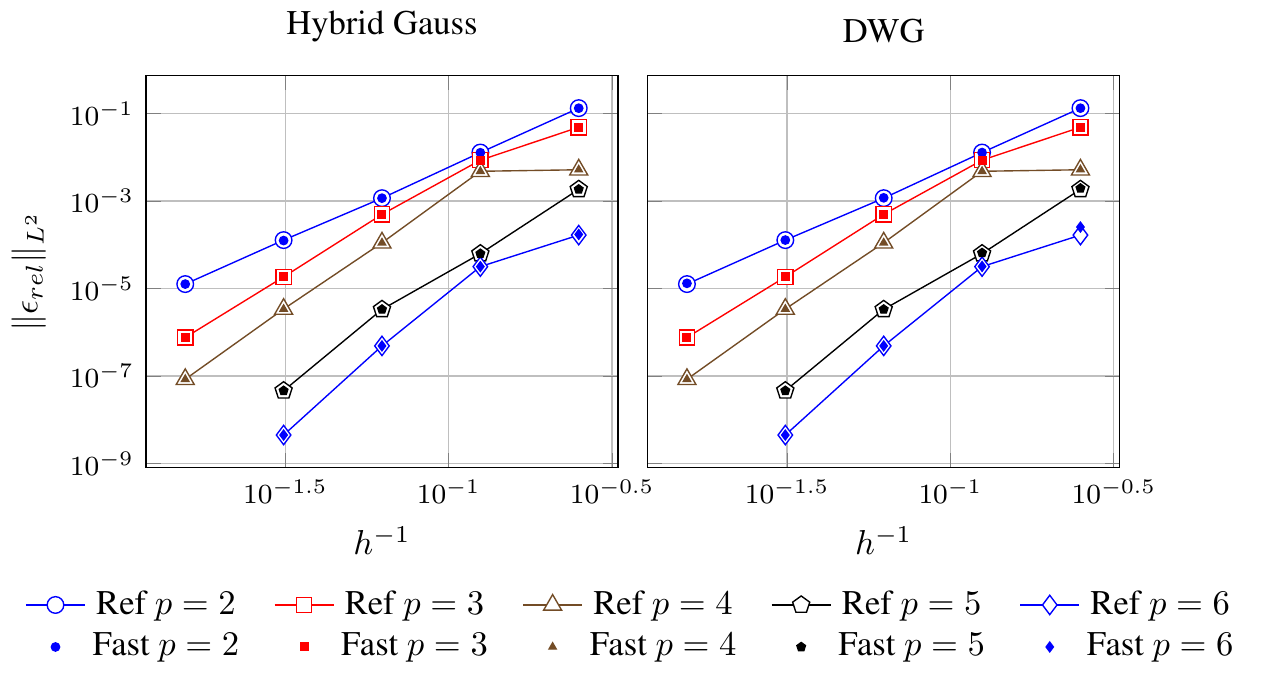}
  \caption{Comparison of the relative approximation error related of the reference solution (solid lines) with approaches using fast assembly and formation (markers): Hybrid Gauss (left figure) and discontinuous weighted quadrature (right figure).}
  \label{fig:exConvergence}
\end{figure}

First, the accuracy of the $L^2$-projection is addressed. Fig.~\ref{fig:exConvergence} summarizes $\| \epsilon_{rel}\|_{L^2}$ for all meshes. In the case of $p=\{5,6\}$, the last refinement step (i.e., $h = 64$)  is missing due to a lack of memory.
In all cases, optimal convergence rates and excellent agreement with the reference solutions can be observed for the Hybrid Gauss and the DWQ approach. 

\begin{figure}[h]
  \centering
  \includegraphics[scale=1.0]{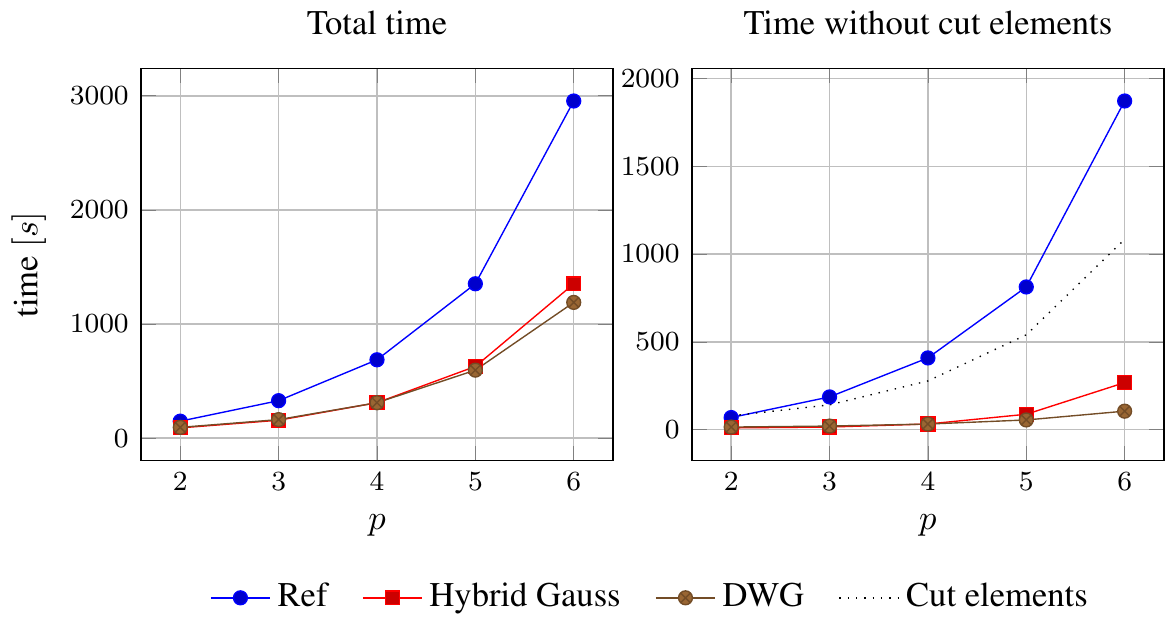}
  \caption{Total timings for setting up the mass matrix using the standard Gauss routines (Ref) and the fast formation and assembly approaches (Hybrid Gauss and DWQ). The figure on the right-hand side reports the time for integrating cut elements (Cut elements) separately since all schemes use the same routines for this task.}
  \label{fig:exTimeTotal}
\end{figure}

Next, Fig.~\ref{fig:exTimeTotal} provides insights into the efficiency of the different approaches by reporting the total timings for setting up the mass matrix for all meshes with $h=32$. 
Looking at the total time, the speed-up of the fast assembly and formation approaches is moderate, i.e., approximately a factor of $2-3$. However, when reporting the integration of cut elements separately (cf., Fig.~\ref{fig:exTimeTotal} on the right), the efficiency improvement of the fast routines becomes more apparent. 
It is worth noting that the employed integration of cut elements focuses on high-order accuracy, not efficiency.
Comparing the graphs related to the Hybrid Gauss and the DWQ approaches, the better scaling w.r.t. the degree of the latter breaks through for p > 4.

\begin{figure}[t]
  \centering
  \includegraphics[scale=0.9]{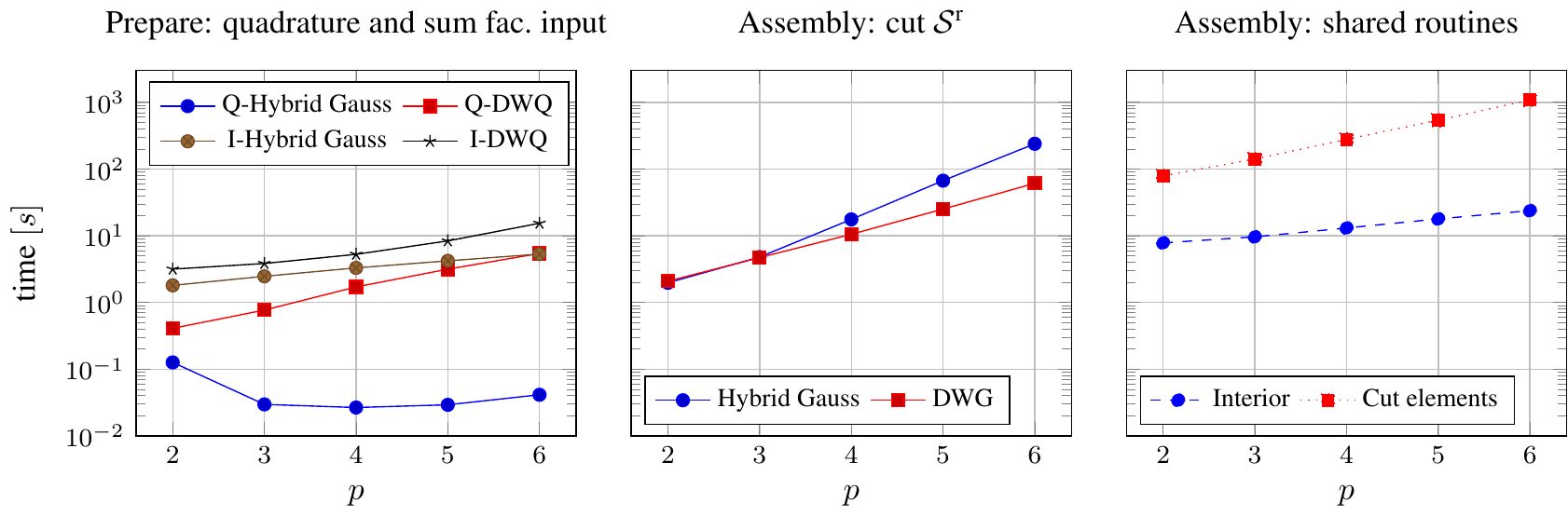}%~a)
  \caption{The various components of the total formation and assembly time for the Hybrid Gauss and DWQ approaches. Left: Preparation time for setting up the weighted quadrature rules (Q) and computing the input for sum factorization (I). Center: time for treating the regular support $\supportdomainReg$ of all cut B-splines. Right: fast formation and assembly for all interior B-splines (Interior) and all cut elements (Cut elements). Note that the timings are provided on a logarithmic scale.}
  \label{fig:exTimePerStep}
\end{figure}

For a more detailed observation, Fig.~\ref{fig:exTimePerStep} lists the different components of the total timing distinguishing between (left) the preparation for weighted quadrature rules and sum factorization, (center) the assembly routine that differs between the Hybrid Gauss and DWQ, and (right) the ones they share.
The left figure indicates the time-critical tasks of DWQ: first, the computation of the DWQ rules, and second, the pre-computation of the quantities at the quadrature points.
The former increases with the number of artificial discontinuities, and the latter growths with the number of nested quadrature points added. 
The figure in the center reveals the advantage of DWQ: the improved efficiency when integrating the regular support $\supportdomainReg$ of all cut B-splines with weighted quadrature rules. For the cases where Hybrid Gauss and DWQ show the same performance (i.e., $p=\{2,3\}$), $\supportdomainReg$ was, in fact, so small that hardly any weighted quadrature was applied. In other words, DWQ reduced to the Hybrid Gauss approach. 
The right figure serves primarily as a reference to compare other contributions to the overall assembly time. 

\section{Conclusion}
A fast formation and assembly approach for fictitious domain methods that use tensor product B-splines as background meshes has been presented.
To be precise, the discontinuous weighted quadrature concept presented in \cite{Marussig2022inproc} has been improved and extended to the three-dimensional setting.
The improvement addresses the layout for the weighted quadrature points. 
Choosing these points as a subset of the Gauss quadrature points allows the reuse of point evaluations and can save the computation of superfluous discontinuous weighted quadrature rules.
Consequently, the discontinuous weighted quadrature performs as well as the Hybrid Gauss approach for moderate degrees and better for higher degrees (in contrast to the results reported in \cite{Marussig2022inproc}).
Still, the computation of multiple weighted quadrature rules is a potential drawback regarding efficiency, which may be addressed simply by parallelization.

\section*{Acknowledgment}
This research was partially supported by the Austrian Science Fund (FWF): M 2806-N, and the joint DFG/FWF Collaborative Research Centre CREATOR (CRC/TRR 361, F90) at TU Darmstadt, TU Graz and JKU Linz.

\bibliographystyle{myplainnat}
\bibliography{FastAssembly4Trim}

\begin{thebibliography}{12}
\providecommand{\natexlab}[1]{#1}
\providecommand{\url}[1]{\texttt{#1}}
\expandafter\ifx\csname urlstyle\endcsname\relax
  \providecommand{\doi}[1]{doi: #1}\else
  \providecommand{\doi}{doi:\begingroup \urlstyle{rm} \Url}\fi

\bibitem[Antolin et~al.(2015)Antolin, Buffa, Calabr\`{o}, Martinelli, and
  Sangalli]{Antolin2015a}
Antolin,~P.; Buffa,~A.; Calabr\`{o},~F.; Martinelli,~M.; Sangalli,~G.:
  Efficient matrix computation for tensor-product isogeometric analysis: The
  use of sum factorization, \emph{Computer Methods in Applied Mechanics and
  Engineering}, 285:\penalty0 817--828, 2015, \doi{10.1016/j.cma.2014.12.013}.

\bibitem[Bressan and Takacs(2019)]{Bressan2019a}
Bressan,~A.; Takacs,~S.: Sum factorization techniques in isogeometric analysis,
  \emph{Computer Methods in Applied Mechanics and Engineering}, 352:\penalty0
  437--460, 2019, \doi{10.1016/j.cma.2019.04.031}.

\bibitem[Calabr{\`o} et~al.(2017)Calabr{\`o}, Sangalli, and Tani]{Calabro2017a}
Calabr{\`o},~F.; Sangalli,~G.; Tani,~M.: Fast formation of isogeometric
  {G}alerkin matrices by weighted quadrature, \emph{Computer Methods in Applied
  Mechanics and Engineering}, 316:\penalty0 606--622, 2017,
  \doi{10.1016/j.cma.2016.09.013}.

\bibitem[Fries and Omerovi{\'c}(2016)]{Fries2015a}
Fries,~T.-P.; Omerovi{\'c},~S.: Higher-order accurate integration of implicit
  geometries, \emph{International Journal for Numerical Methods in
  Engineering}, 106\penalty0 (5):\penalty0 323--371, 2016,
  \doi{10.1002/nme.5121}.

\bibitem[Giannelli et~al.(2022)Giannelli, Kandu\u{c}, Martinelli, Sangalli, and
  Tani]{Giannelli2022a}
Giannelli,~C.; Kandu\u{c},~T.; Martinelli,~M.; Sangalli,~G.; Tani,~M.: Weighted
  quadrature for hierarchical {B}-splines, \emph{Computer Methods in Applied
  Mechanics and Engineering}, 400:\penalty0 115465, 2022,
  \doi{10.1016/j.cma.2022.115465}.

\bibitem[Hiemstra et~al.(2017)Hiemstra, Calabr\`{o}, Schillinger, and
  Hughes]{Hiemstra2017a}
Hiemstra,~R.R.; Calabr\`{o},~F.; Schillinger,~D.; Hughes,~T.J.R.: Optimal and
  reduced quadrature rules for tensor product and hierarchically refined
  splines in isogeometric analysis, \emph{Computer Methods in Applied Mechanics
  and Engineering}, 316:\penalty0 966--1004, 2017,
  \doi{10.1016/j.cma.2016.10.049}.

\bibitem[H{\"o}llig(2003)]{Hoellig2003b}
H{\"o}llig,~K.: \emph{Finite Element Methods with {B}-Splines}, volume~26 of
  \emph{Frontiers in Applied Mathematics}.
\newblock SIAM, 2003.

\bibitem[Marussig(2022)]{Marussig2022inproc}
Marussig,~B.: Fast formation and assembly of isogeometric {G}alerkin matrices
  for trimmed patches.
\newblock In: Manni,~C.; Speleers,~H., editors, \emph{Geometric Challenges in
  Isogeometric Analysis}, pages 149--169, Cham, 2022 Springer International
  Publishing.
\newblock ISBN 978-3-030-92313-6, \doi{10.1007/978-3-030-92313-6_7}.

\bibitem[Marussig and Hughes(2018)]{marussig2017a}
Marussig,~B.; Hughes,~T.J.R.: A review of trimming in isogeometric analysis:
  Challenges, data exchange and simulation aspects, \emph{Archives of
  Computational Methods in Engineering}, 25\penalty0 (4):\penalty0 1059--1127,
  2018, \doi{10.1007/s11831-017-9220-9}.

\bibitem[Marussig et~al.(2016)Marussig, Zechner, Beer, and
  Fries]{marussig2016a}
Marussig,~B.; Zechner,~J.; Beer,~G.; Fries,~T.-P.: Stable isogeometric analysis
  of trimmed geometries, \emph{Computer Methods in Applied Mechanics and
  Engineering}, 316:\penalty0 497--521, 2016, \doi{10.1016/j.cma.2016.07.040}.

\bibitem[Marussig et~al.(2018)Marussig, Hiemstra, and Hughes]{marussig2018a}
Marussig,~B.; Hiemstra,~R.; Hughes,~T.J.R.: Improved conditioning of
  isogeometric analysis matrices for trimmed geometries, \emph{Computer Methods
  in Applied Mechanics and Engineering}, 334:\penalty0 79--110, 2018,
  \doi{10.1016/j.cma.2018.01.052}.

\bibitem[Orszag(1980)]{Orszag1980a}
Orszag,~S.A.: Spectral methods for problems in complex geometries,
  \emph{Journal of Computational Physics}, 37\penalty0 (1):\penalty0 70--92,
  1980, \doi{10.1016/0021-9991(80)90005-4}.

\end{thebibliography}

\end{document}